\def\year{2022}\relax
\documentclass[letterpaper]{article} 
\usepackage{aaai22}  
\usepackage{times}  
\usepackage{helvet}  
\usepackage{courier}  
\usepackage[hyphens]{url}  
\usepackage{graphicx} 
\def\UrlFont{\rm}  
\usepackage{natbib}  
\usepackage{caption} 
\DeclareCaptionStyle{ruled}{labelfont=normalfont,labelsep=colon,strut=off} 
\usepackage{algorithm}
\usepackage{algorithmic}

\usepackage{newfloat}
\usepackage{listings}
\floatstyle{ruled}
\newfloat{listing}{tb}{lst}{}
\floatname{listing}{Listing}
\usepackage[utf8]{inputenc} 
\usepackage{booktabs}       
\usepackage{amsfonts}       
\usepackage{nicefrac}       
\usepackage{microtype}      
\usepackage{xcolor}         

\usepackage{tabularx}
\usepackage{multicol}
\usepackage{multirow}
\usepackage{amsmath}
\usepackage{mathrsfs}
\usepackage{amsthm}
\usepackage{stackengine}
\usepackage{algorithm}
\usepackage{algorithmic}
\usepackage{subcaption}
\newcolumntype{Y}{>{\centering\arraybackslash}X}

\title{AAAI Press Formatting Instructions \\for Authors Using \LaTeX{} --- A Guide}

\title{Unsupervised Domain Adaptation for Constraining Star Formation Histories}
\author {
    Sankalp Gilda,\footnote{Equal Contribution}\textsuperscript{\rm 1}
    Antoine de Mathelin, \footnotemark[1]\textsuperscript{\rm 2}
    Sabine Bellstedt, \textsuperscript{\rm 3}
    Guillaume Richard \textsuperscript{\rm 4}
}
\affiliations {
    \textsuperscript{\rm 1} Fermata Energy\\
    \textsuperscript{\rm 2} Michelin - ENS Paris-Saclay, Centre Borelli\\
    \textsuperscript{\rm 3} International Centre for Radio Astronomy Research, The University of Western Australia \\
    \textsuperscript{\rm 4} EDF R\&D - ENS Paris-Saclay, Centre Borelli \\
    sankalp.gilda@gmail.com, antoine.de-mathelin-de-papigny@michelin.com, sabine.bellstedt@uwa.edu.au, guillaume-gu.richard@edf.fr
}

\usepackage{bibentry}

\begin{document}

\maketitle

\begin{abstract}
The prevalent paradigm of machine learning today is to use past observations to predict future ones. What if, however, we are interested in knowing the past given the present? This situation is indeed one that astronomers must contend with often. To understand the formation of our universe, we must derive the time evolution of the visible mass content of galaxies. However, to observe a complete star life, one would need to wait for one billion years! To overcome this difficulty, astrophysicists leverage supercomputers and evolve simulated models of galaxies till the current age of the universe, thus establishing a mapping between observed radiation and star formation histories (SFHs). Such ground-truth SFHs are lacking for actual galaxy observations, where they are usually inferred -- with often poor confidence -- from spectral energy distributions (SEDs) using Bayesian fitting methods. In this investigation, we discuss the ability of unsupervised domain adaptation to derive accurate SFHs for galaxies with simulated data as a necessary first step in developing a technique that can ultimately be applied to observational data. The code and the data used for the experiments conducted in this work are available on Github\footnote{\url{https://github.com/antoinedemathelin/Unsupervised-Domain-Adaptation-For-Star-Formation-History}}.
\end{abstract}

\section{Introduction}
\label{sec:intro}

In recent times, many transfer problems have arisen, and various methods exist to solve them. We can broadly classify these into three categories. First, supervised domain adaptation, where only a few labeled target data are available \cite{Dai2007TrAdaBoost, de2020adversarial, Motiian2017FewShotAdversarial, Motiian2017UDDA}. Second, semi-supervised domain adaptation, where in addition to these labeled data, a large amount of unlabeled data is available \cite{kumar2010semi, Saito2019SSDA, Tzeng2015SDANN}. Finally, unsupervised domain adaptation, where only unlabeled data is available in the target \cite{Ganin2016DANN, Huang2007KMM, Richard2020MSDA, Saito2018MCD, Sugiyama2007KLIEP}. Researchers have shown that adding target labels can increase the performance of the model \cite{Motiian2017FewShotAdversarial}. This setting is also the most encountered in practice: \cite{Cortes2019GeneralDisc}, as it is often possible to label at least a few target samples.

However, there are cases in which getting supervised data is not possible. Such is the case of the present astrophysics problem of SFH prediction. Here, we aim to learn the history of our Universe through the evolution of its galaxies' masses throughout their individual histories. We try to obtain this history from the radiation that reaches us on earth -- the Spectral Energy Distribution (SED), a function of the brightness of a galaxy with the wavelength of observation.

\begin{figure}
    \centering
    \includegraphics[width=\columnwidth]{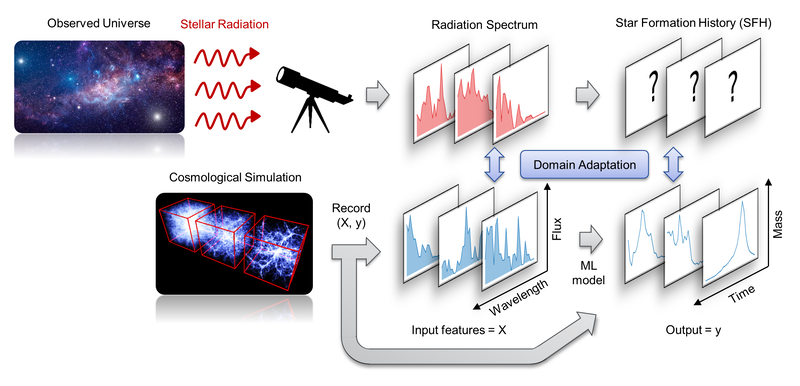}
    \caption{Radiation spectrums with corresponding star formation histories can be recorded from cosmological simulation to form a dataset $(X, y)$. This dataset is used to learn a machine learning model. Because of the domain shift between real radiation spectrums and simulated ones, domain adaptation is used to adapt the learned machine learning model to the real observations.}
    \label{fig:intro}
\end{figure}

As galaxies evolve over billions of years, it is impossible to completely capture any single galaxy's SED as a function of time. Astrophysicists then generate models of light through physically-motivated mechanisms, which allow building artificial star formation histories with the corresponding radiation. One of the objectives of these artificial data sets is to find the correspondence between the radiation and the SFH. We try to find a function that can robustly perform this mapping. Classically, Bayesian models have been developed : \textsc{ProSpect} \cite{Robotham20}, \textsc{MagPhys} \cite{daCunha08}, \textsc{CIGALE} \cite{Noll09}, \textsc{BAGPIPES} \cite{Carnall18a}, \textsc{Prospector} \citep{Johnson17}, and numerous others, more recently \textsc{mirkwood}, \cite{Gilda21} showed that deep learning approaches have the potential to infer SFHs from SEDs.

A significant challenge with applying all the above models arises due to the `domain shift' between simulated and real galaxies. Even a deep neural network (DNN) trained only on the synthetic data sets will falter when predicting the SFHs of real galaxies. Therefore, in this work, we investigate the use of unsupervised domain adaptation as an alternative method of extracting SFHs for individual galaxies without being limited by the simplifications of physical models and parametrizations of conventional SED-fitting techniques. We turn toward cosmological, hydrodynamic simulations, including \textsc{eagle} \citep{Schaye15}, \textsc{illustristng} \citep{Nelson18, Pillepich18} and \textsc{simba} \citep{dave_simba}. These simulate a volume of the Universe from shortly after the Big Bang to the present day, recording the evolution of dark and baryonic matter over time. Galaxies formed within simulations can be studied as a proxy for real galaxies and can be compared directly to observations. They can thus serve to augment our understanding of the evolution of the Universe.

\begin{figure*}
    \centering
    \includegraphics[width=\textwidth]{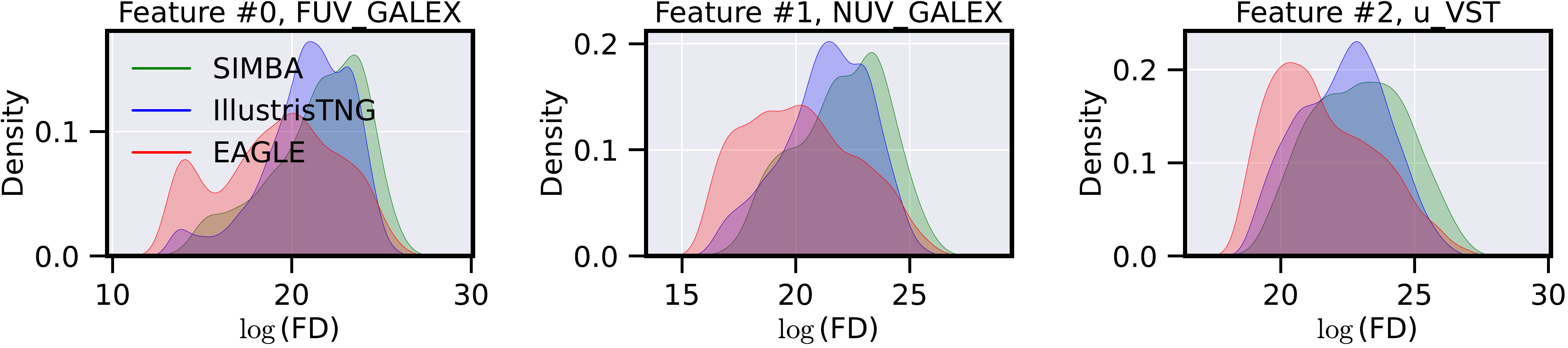}
    \caption{Kernel density estimate plots of the $\log$ of flux densities (FD) for the first three features. The feature names at the top of each plot are the names of the filters, each centered at a different wavelength, in which photometry was simulated. We notice that for the three features shown here, all three simulations share the same support, which justifies our decision of using KLIEP.}
    \label{fig:flux_distribution}
\end{figure*}

\section{Data}\label{sec:data}

Our training and test data sets consist of SEDs from three state-of-the-art cosmological galaxy formation simulations, where the true physical properties are known -- including their true star formation histories. {\sc Simba} \citep{dave_simba}, {\sc Eagle} \citep{schaye_2015_eagle,schaller_2015_eagle,mcalpine_2016_eagle_cat}, and {\sc IllustrisTNG} \citep{vogelsberger2014introducingillustris}, with 1,688, 4,697, and 9,633 samples respectively, together constitute a diverse sample of galaxies with realistic growth histories. Specifically, we select galaxies at a redshift of 0 -- this corresponds to simulated galaxies at the "present" epoch. Each SED is comprised of 20 measurements of the ``brightness" of a galaxy at different wavelength, in the form of flux density (with units of Jansky) and are the co-variates or features that we train/learn on. The outputs/labels are star formation history (SFH) time series vectors with 29 scalar elements, in units of solar mass per year (M$_\odot$ yr$^{-1}$). 

The SFH for a galaxy informs us about the net stellar mass generated as a function of time -- sum of masses of all stars born, less the sum of stellar mass lost in stellar winds as stars age and eventually die. In other words, SFHs are plots of the star formation rates (SFRs) of galaxies against the \emph{lookback time}, which for galaxies at $z=0$ extends from 0 to the very age of our Universe ($\sim$ 13.8 Gyrs). By stacking (adding) together SFHs of several thousand galaxies from a simulation, and dividing by the volume of the simulated box (i.e. the size of the simulated universe), we can derive the cosmic star formation rate density (CSFRD), which is a well-studied global property of our Universe. Astrophysicists aim to tune various physical properties in any given simulation until the the CSFRD plot derived from it matches closely to the most-widely accepted one derived from observations \citep{Madau14}.\footnote{The observed CSFRD (especially at $z>2$) is still a topic of active research currently, as the amount of star formation obscured by dust in the early Universe is still unknown.}

We sequentially train on any two of these simulated data sets, and predict on the third, thus giving us three sets of source- and target-domain data and results.

\section{Methodology}\label{sec:method}

We consider the problem of prediction of star formation history (SFH) where the learner has access to a data set $X \in \mathbb{R}^{n \times p}$ encoding the radiations of $n$ galaxies with $p$ the number of filters/wavelength-bins, and a data set $Y \in \mathbb{R}^{n \times T}$ giving the corresponding SFH for each galaxy with $T$ is the time length of the history. Each SED (row of $X$) is comprised of $p = 20$ measurements of the ``brightness" of a galaxy at different wavelength, in the form of flux density (with units of Jansky) (see Figure \ref{fig:flux_distribution}). Below are our pre-processing steps:
\begin{enumerate}
    \item First, we create three sets of experiments. For each, we use two galaxies in the training and validation sets (with a 9:1 split) and the third galaxy in the test set.
    \item Second, we normalize each SFH time series (each row of $Y$) by its sum and store the resultant normalized SFH (SFH$_{\rm norm}$) and the sum (SFH$_{\rm sum}$) separately. This step is needed because of the large dynamic range of the various star formation histories (see Appendix \ref{sec:best-worse}): SFH curves have a large variety of scales (some increasing to more than $100$ whereas others never increase over $0.1$. By scaling each SFH time series, we make the learning of the curve trend easier. The learning of the SFHs is now decoupled in the learning of SFH$_{\rm norm}$ and SFH$_{\rm sum}$.
    \item Third, we further ease the learning of the SFH$_{\rm norm}$ by reducing the curves to their first $3$ Kernel-PCA components \cite{soentpiet1999advances}. The choice of this decomposition method is motivated by extensive experimentation that showed that Kernel-PCA beat both linear PCA \cite{pca} and discrete wavelet transform \cite{dwt} in their ability to recreate the original time series successfully. For each of the three experiments, we provide the Kernel-PCA with a wide range of hyperparameters and pick the ones that can recreate the original SFH time series back, judged according to the DILATE loss metric \cite{leguen2019} (see Appendix \ref{sec:reduction} and Figure \ref{fig:pca}). The DILATE similarity metric between two time series is defined as an equally weighted average of the dynamic time warping (DTW) and the temporal distortion index (TDI) similarity scores. The number of principal components (3) was chosen by selecting the smallest set that explains at least $80\%$ variance in the validation set. We refer to these kernel-PCA components as SFH$_{\rm kPCA}$.
    \item Fourth, we normalize the input features (the columns from $X$) via log-scaling, and follow this up by standard scaling normalization. In Figure \ref{fig:flux_distribution} we visualize the log-scaled flux densities, in units of Janskies, for all three simulations, in 3 out of 20 filters.
    \item Finally, we derive KLIEP weights for the training samples in all three experiments.
\end{enumerate}

After performing these steps, we apply a domain adaptation method to correct the shift between the source and target input distributions. We choose the method KLIEP \citep{Sugiyama2007KLIEP}, an instance-based method that reweights the sources in order to minimize the KL-divergence between the two domains. Instance-based approaches have been widely used to handle regression domain adaptation issues \cite{Cortes2014DAregression, Huang2007KMM, Mansour2009DATheory, Sugiyama2007KLIEP}, and are particularly robust to negative transfer \cite{de2020adversarial}. We also visually observe on the marginal distributions for the 20 input filters that all domains have the same support in the feature space, which is the framework considered by KLIEP. Finally, KLIEP has the critical advantage of proposing an unsupervised selection procedure to select a relevant bandwidth.

For each of SFH$_{\rm kPCA}$ and SFH$_{\rm sum}$, we train a 4 layer feed-forward DNN with drop-out and 256 nodes in each layer. We use ReLu as the activation function, and Adam \citep{Kingma2014Adam} as the optimizer, with learning rate of $1e^{-3}$. We train for 200 epochs, with early stopping to prevent over-fitting.

\section{Results}\label{sec:results}

There are two main ways of assessing the SFH outputs derived via our machine learning implementation:
\begin{enumerate}
    \item Comparing the derived SFH to the true SFH for individual galaxies in the test set.
    \item Comparing the \textit{predicted} $\Sigma$SFH to the \textit{true} $\Sigma$SFH for each simulation. \textit{True} $\Sigma$SFH is the sum of SFH for all galaxies in a simulation and is a critical metric enabling us to verify the correctness of input physics in a simulation. We know from observations \citep{Madau14} that star formation in the observable Universe peaked about 2 Gyrs after its formation, and all hydrodynamical simulations must produce galaxies that satisfy this observation. By ensuring that the $\Sigma$SFH curve inferred from our neural networks matches those from the underlying simulations, we ground our predictions in science while simultaneously enabling appropriate tuning of model architectures, loss functions, and hyperparameters in case of mismatches. Such comparison also enables us to assess any systematic effects in modeling when training using one simulation and comparing predictions on another.
\end{enumerate}

\begin{figure*}
\begin{subfigure}{0.33\textwidth}
    \centering
    \includegraphics[width=.98\linewidth]{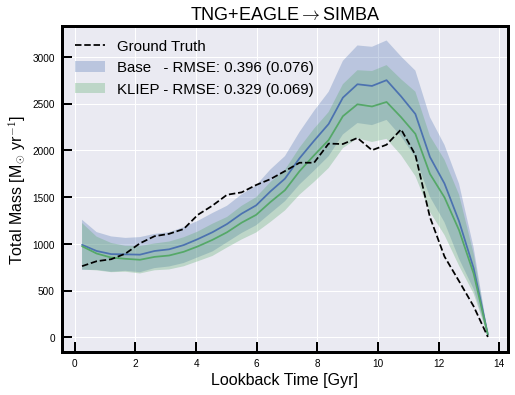}
    \caption{}
    \label{fig:simba_sfh_global_tng}
\end{subfigure}
\hfill
\begin{subfigure}{0.33\textwidth}
    \centering
    \includegraphics[width=.98\linewidth]{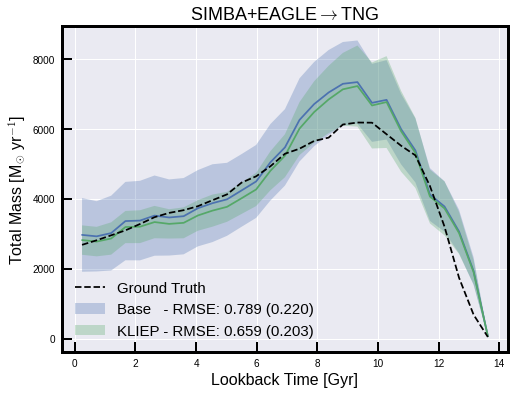}
    \caption{}
    \label{fig:tng_sfh_global}
\end{subfigure}
\hfill
\begin{subfigure}{0.33\textwidth}
    \centering
    \includegraphics[width=.98\linewidth]{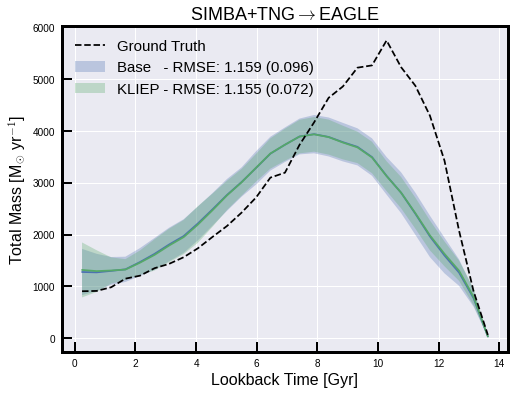}
    \caption{}
    \label{fig:eagle_sfh_global}
\end{subfigure}
\caption{Global SFH predictions for the three experiments. The curves correspond to the sums over all SFH or predicted SFH.}\label{fig:sfh_global}
\end{figure*}

In Figure \ref{fig:sfh_global}, we plot the true and derived $\Sigma$SFH curves for the three distinct test datasets ({\sc Simba}, {\sc IllustrisTNG}, and {\sc Eagle}), where the training data consists of samples from the other two simulations. In these cases, the resulting sample SFHs (and thus $\Sigma$SFH) are most deviant from their ground truth vectors. This is unsurprising, as the different simulations have intrinsically different sample SFHs; this is a known difference between different simulations (as shown in Figure A1 of \citealt{Bellstedt20b}). Tables \ref{tab:metrics} and \ref{tab:metrics_global} show the five metrics tested within this work (MAE, RMSE, BE, DTW, and TDI) for average predictions of SHF, and for $\Sigma$SFH, respectively. Figures \ref{fig:simba_minmax}, \ref{fig:tng_minmax}, and \ref{fig:eagle_minmax} show examples of true and derived SFHs for individual galaxies within the simulations using KLIEP. Based on the five metrics, each row shows an example of the best-performing galaxy output on the left and the worst-performing galaxy on the right. The first thing to note here is that in most cases, the best- and worst-performing galaxies are different when we use different metrics to pick them. This discrepancy highlights the fact that each metric compares time-series differently, and hence it is difficult to pick one metric as the loss function to minimize. 

Another observation made from these three figures is how well our technique can reproduce the stochastic nature of the simulated galaxies' SFHs. As star formation can be an incredibly stochastic process (as is clear from examples such as the top-left panel in Figure \ref{fig:simba_minmax} and the top-left panel in Figure \ref{fig:tng_minmax}), SFHs can regularly fluctuate between high and low values. In general, we find that such stochastic SFHs are poorly recovered. For the sake of galaxy property analysis, the accurate recovery of overall SFH trends is more important than the recovery of individual star formation rate (SFR) epochs. In the bottom-right panel of Figure \ref{fig:simba_minmax}, for example (galaxy index 333) it can be seen that there is a star formation event early on at $\sim 10-12$ Gyr, and then a secondary star formation event from $\sim 2$ Gyr to the present day. Recovering these two main epochs is more crucial than correctly recovering the individual SFR peaks within each epoch. 
One way of achieving this is to temporally smooth the SFHs of individual galaxies prior to training and testing. We can justify such smoothing of simulated features given that we would never expect the derived SFHs for observed galaxies (the ultimate aim of this work) to reproduce such short-scale stochastic features. As an example of this, see Figure C1 of \citet{Robotham20}, where ``good" fits to the SFHs from the semi-analytic model \textsc{Shark} \citep{Lagos18} from the SED-fitting code \textsc{ProSpect} do not recover the stochastic SFHs. 

\begin{table*}
    \centering
    \begin{tabular}{ccccccc}
        \toprule
        \toprule
        Test Simulation &   & RMSE ($\downarrow$) & MAE ($\downarrow$) & BE ($\downarrow$) & DTW ($\downarrow$) & TDI ($\downarrow$) \\
        \midrule
        & Baseline     & $0.3\substack{+0.03 \\ -0.01}$  & $0.23\substack{+0.02 \\ -0.01}$ & $0.04\substack{+0.05 \\ -0.04}$ & $1.0\substack{+0.08 \\ -0.03}$  & $5.13\substack{+0.87 \\ -0.89}$ \\
        \textsc{illustristng} &   \textbf{UDA}   & $\mathbf{0.27}\substack{+0.01 \\ -0.01}$ & $\mathbf{0.2}\substack{+0.01 \\ -0.01}$  & $\mathbf{0.02}\substack{+0.03 \\ -0.02}$ & $\mathbf{0.94}\substack{+0.02 \\ -0.02}$ & $\mathbf{3.19}\substack{+0.82 \\ -0.41}$ \\
        \midrule
        & Baseline      & $0.43\substack{+0.01 \\ -0.01}$ & $0.3\substack{+0.01 \\ -0.01}$  & $-0.08\substack{+0.01 \\ -0.02}$ & $1.59\substack{+0.09 \\ -0.07}$ & $2.91\substack{+2.49 \\ -0.64}$ \\
        \textsc{eagle} & UDA    & $0.43\substack{+0.01 \\ -0.01}$ & $0.3\substack{+0.01 \\ -0.01}$  & $-0.09\substack{+0.01 \\ -0.01}$ & $1.58\substack{+0.07 \\ -0.05}$ & $2.21\substack{+1.08 \\ -0.36}$ \\
        \midrule
        & Baseline    & $0.93\substack{+0.03 \\ -0.02}$ & $0.63\substack{+0.03 \\ -0.02}$ & $0.09\substack{+0.05 \\ -0.03}$  & $3.14\substack{+0.13 \\ -0.11}$ & $4.6\substack{+0.38 \\ -0.94}$  \\
        \textsc{simba} & UDA     & $0.92\substack{+0.02 \\ -0.02}$ & $0.6\substack{+0.02 \\ -0.02}$  & $0.02\substack{+0.03 \\ -0.02}$  & $3.25\substack{+0.1 \\ -0.12}$  & $3.63\substack{+0.88 \\ -0.95}$ \\
        \bottomrule
    \end{tabular}
    \caption{Forecasting results for individual SFH. The first column gives the target domain : {\sc IllustrisTNG}, {\sc Eagle} or {\sc Simba}, the two other domains are used as source domains. The computed metrics between each individual SFH prediction and the corresponding groundtruth are averaged in the target domain. $16^{\rm th}$, $50^{\rm th}$, and $84^{\rm th}$ quantile values are drawn from the 50 predictions per galaxy, corresponding to the 50 neural networks used. Leading method between UDA and no-UDA (baseline) are shown in bold.}
    \label{tab:metrics}
\end{table*}

\begin{table*}
    \centering
    \begin{tabular}{ccccccc}
        \toprule
        \toprule
        Test Simulation &   & RMSE ($\downarrow$) & MAE ($\downarrow$) & BE ($\downarrow$) & DTW ($\downarrow$) & TDI ($\downarrow$) \\
        \midrule
        & Baseline      & $0.72\substack{+0.34 \\ -0.13}$ & $0.61\substack{+0.26 \\ -0.17}$ & $0.34\substack{+0.49 \\ -0.34}$ & $2.64\substack{+1.37 \\ -1.08}$  & $0.17\substack{+0.37 \\ -0.08}$ \\
        \textsc{illustristng} & \textbf{UDA}     & $\mathbf{0.63}\substack{+0.2 \\ -0.15}$  & $\mathbf{0.49}\substack{+0.15 \\ -0.13}$ & $\mathbf{0.21}\substack{+0.25 \\ -0.18}$ & $\mathbf{2.09}\substack{+1.17 \\ -0.64}$  & $\mathbf{0.12}\substack{+0.08 \\ -0.05}$ \\
        \midrule
        & Baseline      & $1.16\substack{+0.1 \\ -0.07}$  & $0.84\substack{+0.11 \\ -0.06}$ & $-0.4\substack{+0.05 \\ -0.08}$  & $3.58\substack{+0.48 \\ -0.5}$  & $0.47\substack{+0.09 \\ -0.11}$ \\
        \textsc{eagle} & UDA   & $1.15\substack{+0.08 \\ -0.08}$ & $0.85\substack{+0.07 \\ -0.09}$ & $-0.4\substack{+0.05 \\ -0.04}$  & $3.52\substack{+0.36 \\ -0.32}$ & $0.47\substack{+0.17 \\ -0.15}$ \\
        \midrule
        & Baseline      & $0.38\substack{+0.09 \\ -0.07}$ & $0.3\substack{+0.06 \\ -0.05}$  & $0.14\substack{+0.09 \\ -0.04}$  & $1.17\substack{+0.39 \\ -0.27}$ & $0.29\substack{+0.1 \\ -0.07}$  \\
        \textsc{simba} &  UDA     & $0.34\substack{+0.05 \\ -0.08}$ & $0.29\substack{+0.05 \\ -0.08}$ & $0.04\substack{+0.04 \\ -0.04}$  & $0.82\substack{+0.22 \\ -0.14}$ & $0.34\substack{+0.16 \\ -0.13}$ \\
        \bottomrule
    \end{tabular}
    \caption{Forecasting results for $\Sigma$SFH (total star formation history). The first column gives the target domain : {\sc IllustrisTNG}, {\sc Eagle} or {\sc Simba}, the two other domains are used as source domains. Predictions for all samples in the target domain are summed and compared to the true $\Sigma$SFH according to the different metrics. $16^{\rm th}$, $50^{\rm th}$, and $84^{\rm th}$ quantile values are drawn from the 50 predictions per galaxy, corresponding to the 50 neural networks used. Leading method between UDA and no-UDA (baseline) are shown in bold. Metrics are scaled (MAE $\times$ 1000, RMSE $\times$ 1000, BE $\times$ 1000, DTW $\times$ 1000 for readability.}
    \label{tab:metrics_global}
\end{table*}

\section{Future Work}\label{sec:future_work}
Results presented here are only one part of a continuing effort to apply sophisticated data-driven methods to SED-fitting, a crucial first step for extracting galaxy properties. 
In future work, we will explore advanced multi-source domain adaptation approaches \citep{Zhao2018MultiSourceDANN, richard2021} which should more robustly account for the presence of multiple source-domain datasets as well as adaptively reweight the training samples depending on the prediction task, thus providing higher quality results.

We are also actively analyzing and attempting to correct systematic biases induced by our various design choices. How do our SFH inferences for galaxies of low mass compare to those of high mass?  Is our modeling sufficient to accurately infer SFH from both old and young galaxies, unlike traditional parametric approaches which suffer at earlier epochs in the Universe's history? Answers to such questions will enable us to discover and correct potential biases in our predictions.

Finally, we also plan on vastly increasing the size of our simulated data to account for the immense diversity in the physics of galaxy formation and evolution modeling--different initial mass functions (IMFs) \citep[for example][]{Salpeter55, Kroupa01, Chabrier03}, nebular emission models  \citep[for example][]{Bruzual03}, mass-metallicity relationships \citep[such as those presented by][]{Tremonti04, Jimmy15, LaraLopez13b}, among others. While this task has a relatively longer horizon--generating and saving new simulations is a compute-intensive task--it is crucial to undertake before data-driven methods such as the one proposed in this paper become trusted by astronomers.

\bibliography{aaai22_gilda}

\clearpage

\onecolumn

\appendix
\section{Selection of Time Series Reduction Method}\label{sec:reduction}

\begin{figure*}[ht]
    \centering
    \includegraphics[width=\textwidth]{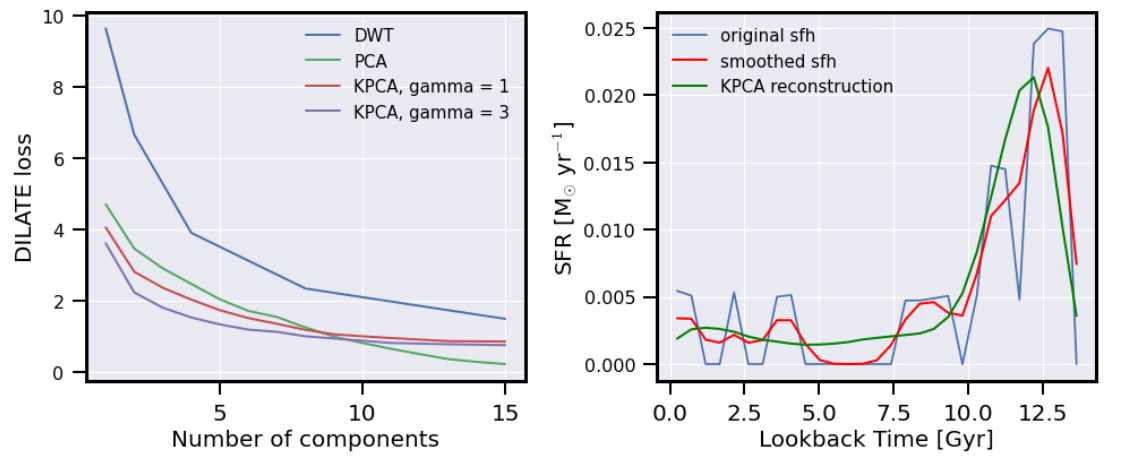}
    \caption{Comparison of SFH reduction approaches for SIMBA. On the left, the evolution of DILATE loss between the true sfh and the reconstructed signal after one of the three transformations: DWT, PCA, kernelPCA. On the right, reconstructed signal for one of the SIMBA SFH.}
    \label{fig:pca}
\end{figure*}

\clearpage

\section{KLIEP Reweighting}

\begin{figure*}[ht]
    \centering
    \includegraphics[width=\textwidth]{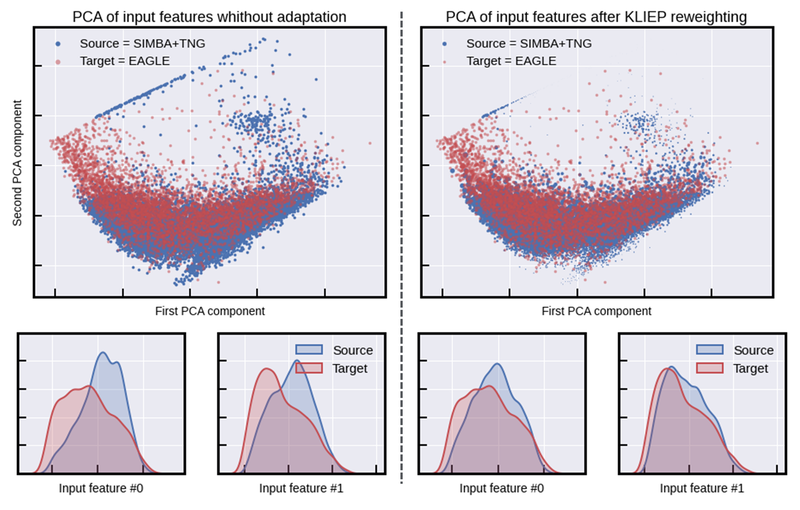}
    \caption{KLIEP brings different domains closer by reweighing source samples. Top: Scatter plot of the first two PCA components in the input space. Bottom: KDE plots of log of the first two features.}
    \label{fig:weights}
\end{figure*}

\clearpage

\section{Extended Results}

\begin{table*}[ht]
    \centering
    \begin{tabular}{ccccccc}
        \toprule
        \toprule
          & RMSE                       & MAE                          & BE                           & DTW                            & TDI                          \\
          \midrule
Base      & $0.43\substack{+0.01 \\ -0.01}$ & $0.3\substack{+0.01 \\ -0.01}$  & $-0.08\substack{+0.01 \\ -0.02}$ & $1.59\substack{+0.09 \\ -0.07}$ & $2.91\substack{+2.49 \\ -0.64}$ \\
KLIEP     & $0.43\substack{+0.01 \\ -0.01}$ & $0.3\substack{+0.01 \\ -0.01}$  & $-0.09\substack{+0.01 \\ -0.01}$ & $1.58\substack{+0.07 \\ -0.05}$ & $2.21\substack{+1.08 \\ -0.36}$ \\
MDD       & $0.45\substack{+0.02 \\ -0.01}$ & $0.32\substack{+0.02 \\ -0.01}$ & $-0.08\substack{+0.02 \\ -0.02}$ & $1.69\substack{+0.12 \\ -0.07}$ & $3.05\substack{+3.04 \\ -0.59}$ \\
DANN      & $0.44\substack{+0.02 \\ -0.01}$ & $0.31\substack{+0.01 \\ -0.01}$ & $-0.08\substack{+0.02 \\ -0.02}$ & $1.64\substack{+0.09 \\ -0.09}$ & $2.9\substack{+3.01 \\ -0.38}$  \\
DeepCORAL & $1.01\substack{+0.08 \\ -0.08}$ & $0.77\substack{+0.07 \\ -0.05}$ & $0.17\substack{+0.06 \\ -0.07}$  & $4.68\substack{+0.42 \\ -0.43}$ & $5.04\substack{+0.29 \\ -0.29}$ \\
KMM       & $0.46\substack{+0.02 \\ -0.02}$ & $0.32\substack{+0.02 \\ -0.01}$ & $-0.09\substack{+0.05 \\ -0.04}$ & $1.73\substack{+0.17 \\ -0.13}$ & $3.42\substack{+2.24 \\ -0.74}$ \\
CORAL     & $0.48\substack{+0.02 \\ -0.02}$ & $0.35\substack{+0.01 \\ -0.02}$ & $0.09\substack{+0.02 \\ -0.02}$  & $1.39\substack{+0.05 \\ -0.08}$ & $2.94\substack{+1.55 \\ -0.41}$ \\
\bottomrule
 \end{tabular}
    \caption{Forecasting results, with {\sc IllustrisTNG} and {\sc Simba} as source domains and {\sc Eagle} as the target domain. Predictions for all samples in {\sc Eagle} have been averaged, and $16^{\rm th}$, $50^{\rm th}$, and $84^{\rm th}$ quantile values are drawn from the 50 predictions per galaxy, corresponding to the 50 neural networks used.}
\end{table*}

\begin{table*}[ht]
    \centering
    \begin{tabular}{ccccccc}
        \toprule
        \toprule
          & RMSE                       & MAE                          & BE                           & DTW                            & TDI                          \\
          \midrule
Base      & $1.16\substack{+0.1 \\ -0.07}$  & $0.84\substack{+0.11 \\ -0.06}$ & $-0.4\substack{+0.05 \\ -0.08}$  & $3.58\substack{+0.48 \\ -0.5}$  & $0.47\substack{+0.09 \\ -0.11}$ \\
KLIEP     & $1.15\substack{+0.08 \\ -0.08}$ & $0.85\substack{+0.07 \\ -0.09}$ & $-0.4\substack{+0.05 \\ -0.04}$  & $3.52\substack{+0.36 \\ -0.32}$ & $0.47\substack{+0.17 \\ -0.15}$ \\
MDD       & $1.3\substack{+0.16 \\ -0.11}$  & $1.03\substack{+0.11 \\ -0.09}$ & $-0.37\substack{+0.09 \\ -0.11}$ & $4.11\substack{+0.57 \\ -0.44}$ & $0.73\substack{+0.2 \\ -0.2}$   \\
DANN      & $1.18\substack{+0.08 \\ -0.13}$ & $0.86\substack{+0.06 \\ -0.08}$ & $-0.4\substack{+0.1 \\ -0.11}$   & $3.59\substack{+0.47 \\ -0.54}$ & $0.46\substack{+0.04 \\ -0.07}$ \\
DeepCORAL & $0.94\substack{+0.23 \\ -0.24}$ & $0.86\substack{+0.24 \\ -0.22}$ & $0.81\substack{+0.29 \\ -0.35}$  & $2.6\substack{+0.65 \\ -0.59}$  & $0.39\substack{+0.13 \\ -0.16}$ \\
KMM       & $1.2\substack{+0.12 \\ -0.12}$  & $0.91\substack{+0.08 \\ -0.13}$ & $-0.4\substack{+0.22 \\ -0.17}$  & $3.77\substack{+0.87 \\ -1.01}$ & $0.56\substack{+0.21 \\ -0.2}$  \\
CORAL     & $0.83\substack{+0.24 \\ -0.1}$  & $0.63\substack{+0.14 \\ -0.08}$ & $0.41\substack{+0.1 \\ -0.1}$    & $1.68\substack{+0.96 \\ -0.41}$ & $0.21\substack{+0.14 \\ -0.04}$ \\
\bottomrule
 \end{tabular}
    \caption{Forecasting results for $\Sigma$SFH (total star formation history) with {\sc IllustrisTNG} and {\sc Simba} as source domains and {\sc Eagle} as the target domain. Predictions for all samples in {\sc Eagle} have been added, and $16^{\rm th}$, $50^{\rm th}$, and $84^{\rm th}$ quantile values are drawn from the 50 predictions per galaxy, corresponding to the 50 neural networks used. Metrics are scaled (MAE $\times$ 1000, RMSE $\times$ 1000, BE $\times$ 1000, DTW $\times$ 1000 for readability.}
\end{table*}

\begin{table*}[ht]
    \centering
    \begin{tabular}{ccccccc}
        \toprule
        \toprule
          & RMSE                       & MAE                          & BE                           & DTW                            & TDI                          \\
          \midrule                          
Base      & $0.93\substack{+0.03 \\ -0.02}$ & $0.63\substack{+0.03 \\ -0.02}$ & $0.09\substack{+0.05 \\ -0.03}$  & $3.14\substack{+0.13 \\ -0.11}$ & $4.6\substack{+0.38 \\ -0.94}$  \\
KLIEP     & $0.92\substack{+0.02 \\ -0.02}$ & $0.6\substack{+0.02 \\ -0.02}$  & $0.02\substack{+0.03 \\ -0.02}$  & $3.25\substack{+0.1 \\ -0.12}$  & $3.63\substack{+0.88 \\ -0.95}$ \\
MDD       & $0.92\substack{+0.02 \\ -0.02}$ & $0.61\substack{+0.02 \\ -0.01}$ & $0.1\substack{+0.05 \\ -0.05}$   & $3.3\substack{+0.14 \\ -0.23}$  & $4.34\substack{+0.49 \\ -0.44}$ \\
DANN      & $0.94\substack{+0.03 \\ -0.02}$ & $0.65\substack{+0.03 \\ -0.02}$ & $0.11\substack{+0.06 \\ -0.04}$  & $3.15\substack{+0.13 \\ -0.14}$ & $4.5\substack{+0.56 \\ -1.0}$   \\
DeepCORAL & $2.19\substack{+0.14 \\ -0.17}$ & $1.74\substack{+0.15 \\ -0.17}$ & $1.06\substack{+0.18 \\ -0.18}$  & $9.81\substack{+0.87 \\ -0.97}$ & $7.55\substack{+0.22 \\ -0.2}$  \\
KMM       & $1.02\substack{+0.04 \\ -0.04}$ & $0.68\substack{+0.04 \\ -0.03}$ & $0.09\substack{+0.09 \\ -0.08}$  & $3.47\substack{+0.31 \\ -0.11}$ & $4.55\substack{+0.76 \\ -0.51}$ \\
CORAL     & $0.95\substack{+0.02 \\ -0.01}$ & $0.58\substack{+0.02 \\ -0.02}$ & $-0.22\substack{+0.03 \\ -0.02}$ & $4.29\substack{+0.09 \\ -0.09}$ & $4.4\substack{+0.39 \\ -0.62}$ \\
\bottomrule
 \end{tabular}
    \caption{Forecasting results, with {\sc IllustrisTNG} and {\sc Eagle} as source domains and {\sc Simba} as the target domain. Predictions for all samples in {\sc Simba} have been averaged, and $16^{\rm th}$, $50^{\rm th}$, and $84^{\rm th}$ quantile values are drawn from the 50 predictions per galaxy, corresponding to the 50 neural networks used.}
\end{table*}

\begin{table*}[ht]
    \centering
    \begin{tabular}{ccccccc}
        \toprule
        \toprule
          & RMSE                       & MAE                          & BE                           & DTW                            & TDI                          \\
          \midrule
Base      & $0.38\substack{+0.09 \\ -0.07}$ & $0.3\substack{+0.06 \\ -0.05}$  & $0.14\substack{+0.09 \\ -0.04}$  & $1.17\substack{+0.39 \\ -0.27}$ & $0.29\substack{+0.1 \\ -0.07}$  \\
KLIEP     & $0.34\substack{+0.05 \\ -0.08}$ & $0.29\substack{+0.05 \\ -0.08}$ & $0.04\substack{+0.04 \\ -0.04}$  & $0.82\substack{+0.22 \\ -0.14}$ & $0.34\substack{+0.16 \\ -0.13}$ \\
MDD       & $0.29\substack{+0.12 \\ -0.07}$ & $0.22\substack{+0.1 \\ -0.05}$  & $0.16\substack{+0.08 \\ -0.08}$  & $0.98\substack{+0.47 \\ -0.28}$ & $0.24\substack{+0.13 \\ -0.08}$ \\
DANN      & $0.42\substack{+0.12 \\ -0.07}$ & $0.33\substack{+0.1 \\ -0.05}$  & $0.19\substack{+0.11 \\ -0.07}$  & $1.38\substack{+0.55 \\ -0.33}$ & $0.28\substack{+0.13 \\ -0.06}$ \\
DeepCORAL & $1.91\substack{+0.31 \\ -0.32}$ & $1.8\substack{+0.3 \\ -0.31}$   & $1.8\substack{+0.3 \\ -0.31}$    & $9.38\substack{+1.8 \\ -1.84}$  & $2.12\substack{+0.4 \\ -0.93}$  \\
KMM       & $0.38\substack{+0.08 \\ -0.11}$ & $0.31\substack{+0.06 \\ -0.09}$ & $0.15\substack{+0.14 \\ -0.14}$  & $1.11\substack{+0.38 \\ -0.43}$ & $0.27\substack{+0.15 \\ -0.09}$ \\
CORAL     & $0.49\substack{+0.05 \\ -0.06}$ & $0.41\substack{+0.04 \\ -0.05}$ & $-0.37\substack{+0.06 \\ -0.03}$ & $2.02\substack{+0.35 \\ -0.38}$ & $0.96\substack{+0.31 \\ -0.23}$ \\
\bottomrule
 \end{tabular}
    \caption{Forecasting results for $\Sigma$SFH (total star formation history) with {\sc IllustrisTNG} and {\sc Eagle} as source domains and {\sc Simba} as the target domain. Predictions for all samples in {\sc Simba} have been added, and $16^{\rm th}$, $50^{\rm th}$, and $84^{\rm th}$ quantile values are drawn from the 50 predictions per galaxy, corresponding to the 50 neural networks used. Metrics are scaled (MAE $\times$ 1000, RMSE $\times$ 1000, BE $\times$ 1000, DTW $\times$ 1000 for readability.}
\end{table*}

\begin{table*}[ht]
    \centering
    \begin{tabular}{ccccccc}
        \toprule
        \toprule
          & RMSE                       & MAE                          & BE                           & DTW                            & TDI                          \\
          \midrule 
Base      & $0.3\substack{+0.03 \\ -0.01}$  & $0.23\substack{+0.02 \\ -0.01}$ & $0.04\substack{+0.05 \\ -0.04}$ & $1.0\substack{+0.08 \\ -0.03}$  & $5.13\substack{+0.87 \\ -0.89}$ \\
KLIEP     & $0.27\substack{+0.01 \\ -0.01}$ & $0.2\substack{+0.01 \\ -0.01}$  & $0.02\substack{+0.03 \\ -0.02}$ & $0.94\substack{+0.02 \\ -0.02}$ & $3.19\substack{+0.82 \\ -0.41}$ \\
MDD       & $0.3\substack{+0.02 \\ -0.01}$  & $0.23\substack{+0.02 \\ -0.01}$ & $0.02\substack{+0.04 \\ -0.03}$ & $0.98\substack{+0.08 \\ -0.04}$ & $4.92\substack{+0.85 \\ -0.58}$ \\
DANN      & $0.3\substack{+0.02 \\ -0.01}$  & $0.22\substack{+0.02 \\ -0.01}$ & $0.02\substack{+0.05 \\ -0.05}$ & $1.01\substack{+0.08 \\ -0.07}$ & $4.99\substack{+0.72 \\ -0.9}$  \\
DeepCORAL & $0.5\substack{+0.04 \\ -0.03}$  & $0.41\substack{+0.04 \\ -0.03}$ & $0.01\substack{+0.08 \\ -0.05}$ & $2.08\substack{+0.25 \\ -0.16}$ & $6.74\substack{+0.68 \\ -0.34}$ \\
KMM       & $0.33\substack{+0.05 \\ -0.03}$ & $0.25\substack{+0.04 \\ -0.03}$ & $0.05\substack{+0.06 \\ -0.04}$ & $1.1\substack{+0.13 \\ -0.08}$  & $4.83\substack{+0.67 \\ -0.75}$ \\
CORAL     & $0.55\substack{+0.25 \\ -0.09}$ & $0.45\substack{+0.21 \\ -0.07}$ & $0.32\substack{+0.12 \\ -0.12}$ & $2.08\substack{+1.3 \\ -0.48}$  & $6.49\substack{+0.88 \\ -1.24}$ \\
\bottomrule
 \end{tabular}
    \caption{Forecasting results, with {\sc Simba} and {\sc Eagle} as source domains and {\sc IllustrisTNG} as the target domain. Predictions for all samples in {\sc IllustrisTNG} have been averaged, and $16^{\rm th}$, $50^{\rm th}$, and $84^{\rm th}$ quantile values are drawn from the 50 predictions per galaxy, corresponding to the 50 neural networks used.}
\end{table*}

\begin{table*}[ht]
    \centering
    \begin{tabular}{ccccccc}
        \toprule
        \toprule
          & RMSE                       & MAE                          & BE                           & DTW                            & TDI                          \\
          \midrule
Base      & $0.72\substack{+0.34 \\ -0.13}$ & $0.61\substack{+0.26 \\ -0.17}$ & $0.34\substack{+0.49 \\ -0.34}$ & $2.64\substack{+1.37 \\ -1.08}$  & $0.17\substack{+0.37 \\ -0.08}$ \\
KLIEP     & $0.63\substack{+0.2 \\ -0.15}$  & $0.49\substack{+0.15 \\ -0.13}$ & $0.21\substack{+0.25 \\ -0.18}$ & $2.09\substack{+1.17 \\ -0.64}$  & $0.12\substack{+0.08 \\ -0.05}$ \\
MDD       & $0.79\substack{+0.36 \\ -0.2}$  & $0.64\substack{+0.26 \\ -0.17}$ & $0.24\substack{+0.42 \\ -0.3}$  & $2.51\substack{+1.55 \\-0.67}$  & $0.15\substack{+0.24 \\ -0.06}$ \\
DANN      & $0.76\substack{+0.27 \\ -0.19}$ & $0.61\substack{+0.23 \\ -0.12}$ & $0.21\substack{+0.43 \\ -0.51}$ & $2.57\substack{+1.01 \\ -0.77}$  & $0.19\substack{+0.4 \\ -0.09}$  \\
DeepCORAL & $0.68\substack{+0.44 \\ -0.17}$ & $0.58\substack{+0.46 \\ -0.2}$  & $0.06\substack{+0.8 \\ -0.51}$  & $1.93\substack{+1.67 \\ -0.66}$  & $0.38\substack{+0.46 \\ -0.28}$ \\
KMM       & $0.85\substack{+0.57 \\ -0.32}$ & $0.67\substack{+0.43 \\ -0.27}$ & $0.44\substack{+0.55 \\ -0.37}$ & $2.87\substack{+2.89 \\ -1.34}$  & $0.16\substack{+0.17 \\ -0.08}$ \\
CORAL     & $3.85\substack{+1.92 \\ -0.8}$  & $3.23\substack{+1.44 \\ -0.74}$ & $3.04\substack{+1.11 \\ -1.14}$ & $18.75\substack{+12.24 \\ -6.1}$ & $0.4\substack{+3.29 \\ -0.38}$ \\
\bottomrule
 \end{tabular}
    \caption{Forecasting results for $\Sigma$SFH (total star formation history) with {\sc Simba} and {\sc Eagle} as source domains and {\sc IllustrisTNG} as the target domain. Predictions for all samples in {\sc IllustrisTNG} have been added, and $16^{\rm th}$, $50^{\rm th}$, and $84^{\rm th}$ quantile values are drawn from the 50 predictions per galaxy, corresponding to the 50 neural networks used. Metrics are scaled (MAE $\times$ 1000, RMSE $\times$ 1000, BE $\times$ 1000, DTW $\times$ 1000 for readability.}
\end{table*}

\clearpage

\begin{figure*}
\begin{subfigure}{\textwidth}
    \centering
    \includegraphics[width=0.7\linewidth]{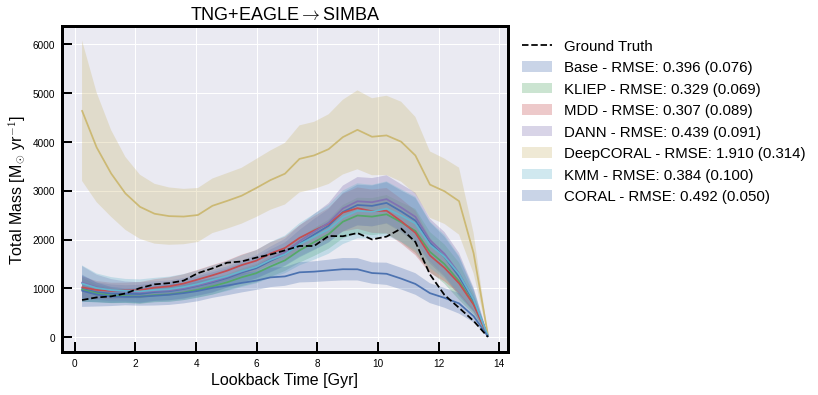}
    \caption{}
\end{subfigure}
\hfill
\begin{subfigure}{\textwidth}
    \centering
    \includegraphics[width=0.7\linewidth]{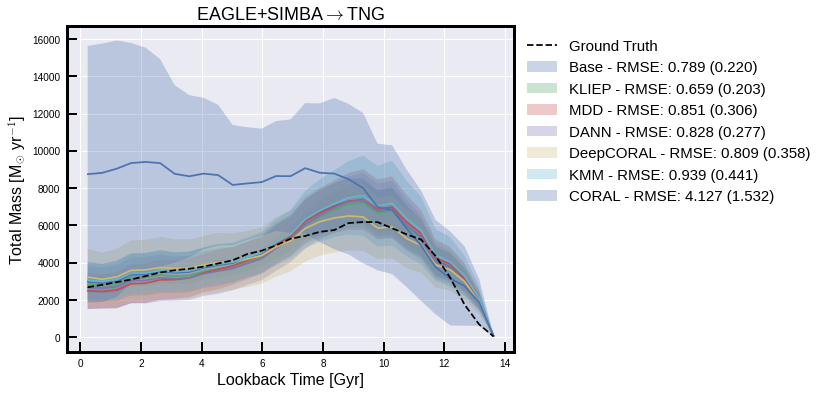}
    \caption{}
\end{subfigure}
\hfill
\begin{subfigure}{\textwidth}
    \centering
    \includegraphics[width=0.7\linewidth]{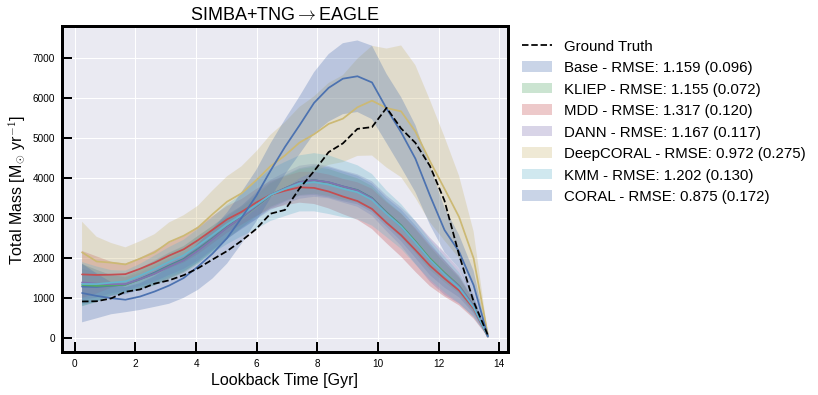}
    \caption{}
\end{subfigure}
\caption{Global SFH predictions for the three experiments. The curves correspond to the sums over all SFH or predicted SFH.}
\end{figure*}

\clearpage

\section{Predictions for individual SFHs}\label{sec:best-worse}

\begin{figure*}[ht]
    \centering
    \includegraphics[width=0.7\textwidth]{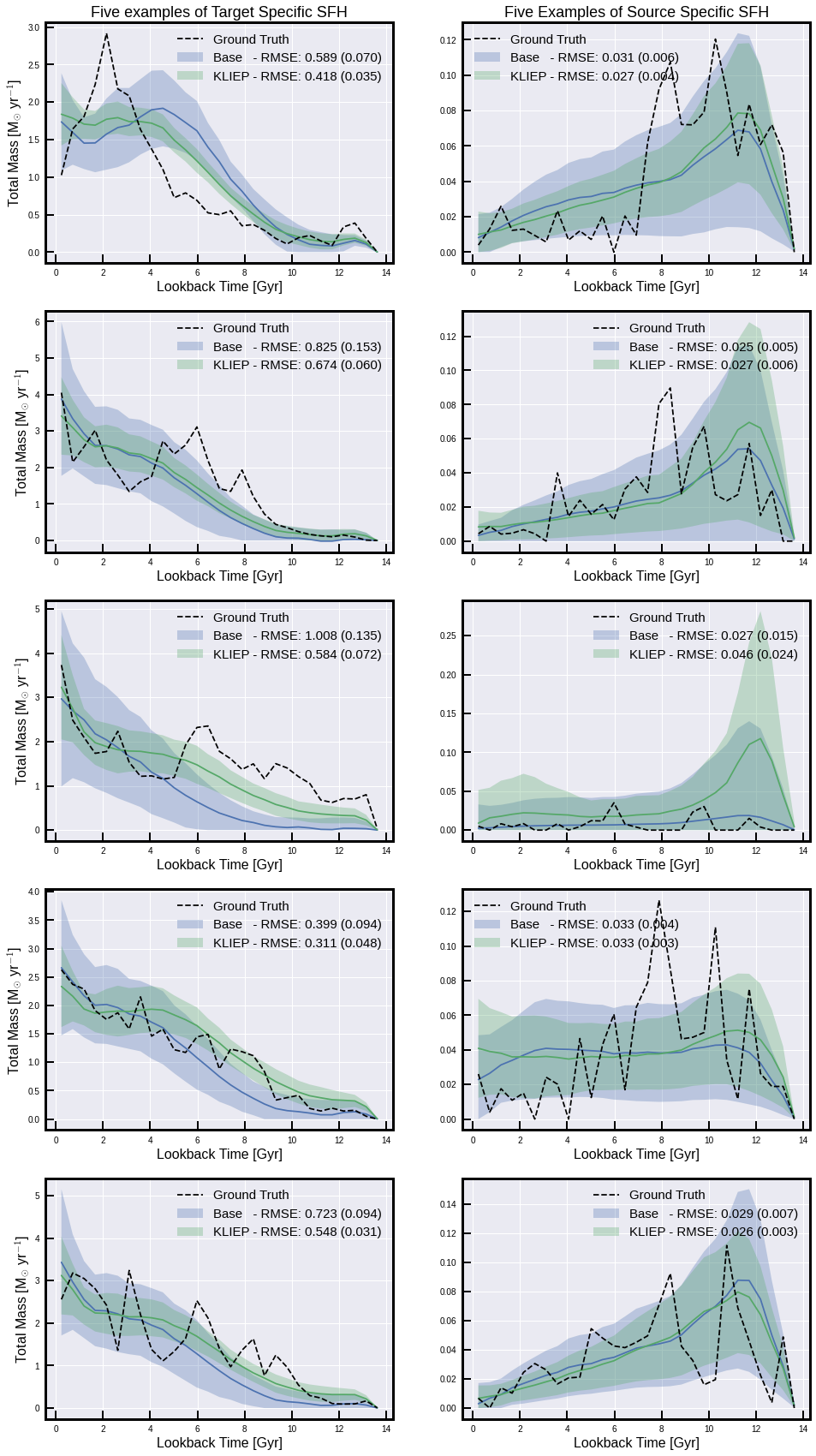}
    \caption{Individuals galaxies for {\sc Simba} from target and source specific clusters.}
    \label{fig:simba_minmax}
\end{figure*}

\begin{figure*}[ht]
    \centering
    \includegraphics[width=0.7\textwidth]{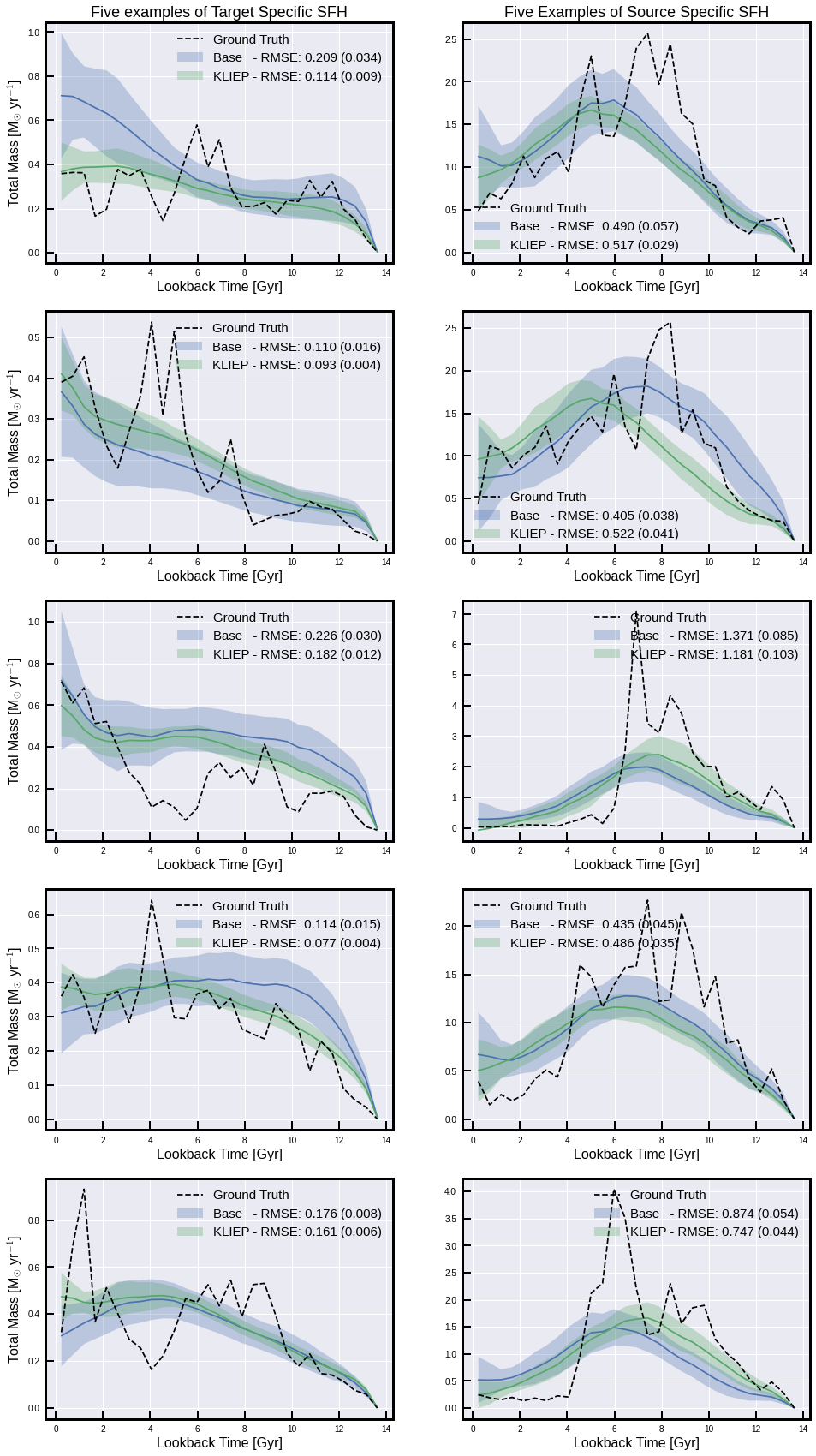}
    \caption{Individuals galaxies for {\sc IllustrisTNG} from target and source specific clusters.}
    \label{fig:tng_minmax}
\end{figure*}

\begin{figure*}[ht]
    \centering
    \includegraphics[width=0.7\textwidth]{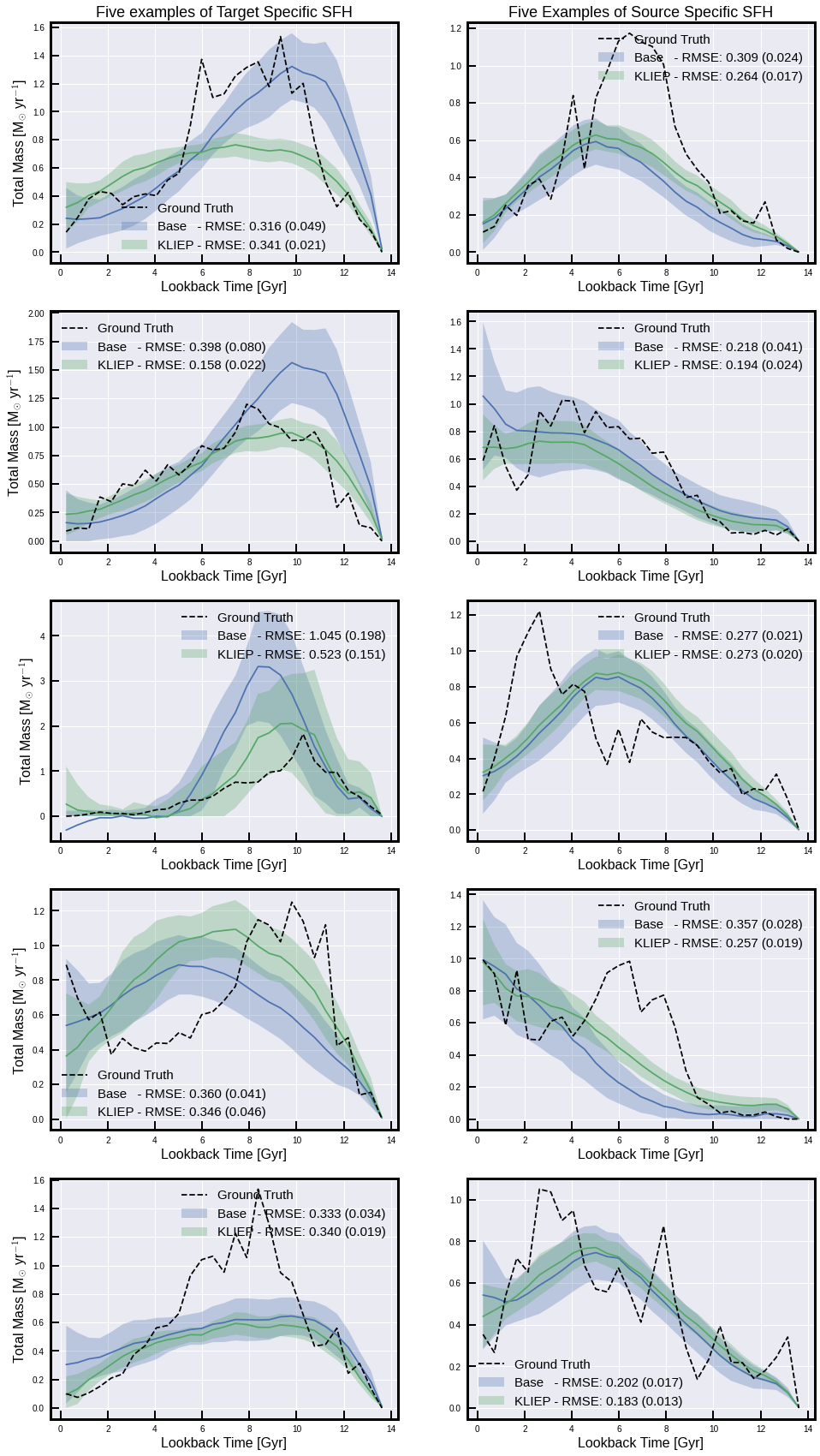}
    \caption{Individuals galaxies for {\sc Eagle} from target and source specific clusters.}
    \label{fig:eagle_minmax}
\end{figure*}

\end{document}